# The Far Ultraviolet Spectroscopic picture of the Missing Baryons in the Local Group

Fabrizio Nicastro*, Andreas Zezas*, Martin Elvis*, Smita Mathur†, Fabrizio Fiore‡, Cesare Cecchi-Pestellini*, Douglas Burke*, Jeremy Drake*, Piergiorgio Casella*

*Harvard-Smithsonian Center for Astrophysics 60 Garden St, Cambridge MA 02138

†Astronomy Department, The Ohio State University, 43210, Columbus, OH, USA

‡Osservatorio Astronomico di Monteporzio, Via Osservatorio Via Frascati 33, Monteporzio-Catone (RM), I-00040 Italy

**The number of detected baryons in the low-redshift Universe ($z \leq 1$) is far smaller than the corresponding number of baryons observed at higher redshift. According to hydrodynamical simulations for the formation of structure in the Universe[1,2,3], up to 2/3 of the 'missing baryons' may have escaped identification so far, due to their high temperature and low density. One of the few ways to detect this matter is to look for its UV absorption signature on the light from background sources. Here we show that the amplitude of the average velocity vector of 'high velocity' OVI absorbers detected in a UV survey of AGN decreases significantly in moving from the Local Standard of Rest, to the Galactic Standard of Rest and Local Group frames. At least 82 % of these absorbers are not associated with any 'high velocity' HI complex (HV-HI) in our Galaxy, and so are strongly implicated as being due to primordial Warm-Hot Intergalactic Medium (WHIM) pervading an extended Galactic Corona or our Local Group. We estimate that the total baryonic mass**

**contained in this medium is as large as ~ $10^{12}$ M , similar to the combined mass of the two most massive virialized constituents of the Local Group (the Andromeda and the Milky Way galaxies), and of the order of the mass required to dynamically stabilize the Local Group[4].**

The first evidence for an absorption system due to the WHIM has recently been discovered in the X-ray and UV spectra of the blazar PKS 2155-304[5] obtained with the Chandra Low Energy Transmission Grating (LETG)[6], and the Far-Ultraviolet Spectroscopic Explorer (FUSE)[7]. Clearly though one line of sight is inadequate to fully define this absorbing medium. FUSE observations of Active Galactic Nuclei (AGN) show two types of ubiquitous z ~ 0 OVI absorbers, similar to those observed in PKS 2155-304: (1) Low Velocity OVI clouds (LV-OVI, $|v_{LSR}|$ <100 km s$^{-1}$)[8,9,10,11,12]; and (2) High Velocity OVI clouds (HV-OVI, $|v_{LSR}|$ >100 km s$^{-1}$)[13,14,15,16]. To investigate the nature of these absorbers we have examined a sample of AGNs using publicly available FUSE data. This has allowed us to clearly identify the HV-OVI component with diffuse gas in an extended Galactic corona or in the Local Group.

The OVI velocity distribution in the Local Standard of Rest (LSR) (Fig. 1, dashed histogram) shows a narrow peak between ±100 km s$^{-1}$ (LV-OVI) with a much broader, roughly symmetric distribution (Fig. 1, solid histogram) extending to ±550 km s$^{-1}$ (HV-OVI). The bimodality of this distribution suggests that LV- and HV-OVI systems belong to two different populations of absorbers, as previously pointed out by Savage et al.[8], and Sembach et al.[13] In the following we concentrate on the HV-OVI absorbers, and use the LV-OVI absorbers only as a comparison sample. A complete discussion of both LV and HV systems is deferred to a forthcoming paper (Nicastro et al., in preparation).

A plot in Galactic coordinates of the LSR velocity distributions for the HV (Fig. 2a, upper panel) absorption systems shows that their LSR velocities split the sky in two distinct halves (Fig. 2a): the hemisphere with $0° \leq l \leq 180°$ contains only HV-OVI lines with **negative** velocities, while the other hemisphere contains mostly HV-OVI absorbers with **positive** LSR velocity. There are three exceptions: two of these have LSR velocities very close to the threshold velocity of $|v_{LSR}| = 100$ km s$^{-1}$, and so may well belong to the LV-OVI population. The third lies at very high latitude, where the concept of longitude becomes meaningless. For comparison, the population of LV-OVI absorbers show no velocity segregation in the LSR.

The red circles in Figure 2 mark the 6 HV-OVI absorbers (~18 % of the sample) which are tentatively identified with 21-cm high velocity HI clouds or complexes, either because of spatial and velocity coincidence (within broad ranges) with entries in the Stark et al.[17] and Wakker et al.[18] catalogs, or because identified as such by Sembach et al.[16] in their HV-OVI compilation (in which ~20 % of HV-OVI are tentatively identified with HV-HI). We stress here that, these are the only 6 HV-OVI absorbers, in either compilation (ours and Sembach's et al.[16]), that may be identified with HV-HI clouds with a known distance, either in our Galaxy's inner halo (with distances less than ~10 kpc), or between us and the two Magellanic Clouds. To be conservative we do not consider further these 6 HV-OVI absorbers.

The systematic LSR velocity distribution of HV-OVI is consistent with matter that is either: (1) counter-rotating, with respect with the Galaxy disk rotation, on orbits external to the sun's orbit, (2) at rest in the Galactic halo or (3) at rest in the intergalactic space surrounding the Galaxy. The range of radial LSR velocities of the HV-OVI ($100 < |v_{LSR}|^{HV} < 550$ km s$^{-1}$) greatly exceeds the range of observed radial velocities in the Galactic disk or halo, suggesting that the Galaxy related options [(1) and (2)] are unlikely.

Moreover some of the HV-OVI velocities exceed a plausible measure of the escape velocity from the Milky way[19]. Finally, clouds in the Galaxy's halo would probably be rotating on random orbits around the galaxy's center, as Globular Clusters do. The peculiar velocities of these clouds along these orbits would tend to randomize the apparent symmetry induced in the LSR by the circular motion of the Sun in the Galaxy for matter effectively at rest in the halo, as observed in Globular Clusters. An intergalactic origin, with distances larger than ~100 kpc, is more consistent with the data.

The LSR is not a rest frame system for the HV-OVI absorbers (Figure 2a, Table 1). This is confirmed by translations of the velocity distribution to other convenient rest frames: the Galactic Standard of Rest (GSR) and the Local Group Standard of Rest (LGSR). The symmetry present in the LSR velocity distribution of the HV-OVI systems (Fig. 2a) disappears in the LGSR (Fig. 2b) and instead appears random, suggesting that the LGSR is a privileged reference frame for the HV-OVI absorbers. The amplitude of the HV-OVI average velocity vector decreases monotonically from the LSR to the GSR, to the LGSR, at which point it is only $<|v_{LGSR}|^{HV}> = 32$ km s$^{-1}$ (Table 1), much smaller than the corresponding velocities in the LSR, and close to the FUSE resolution (~ 20 km s$^{-1}$ at 1032 Å). At the same time, the direction of the average vector becomes very poorly constrained (Table 1). Both results suggest that GSR and LGSR are preferred rest frame for the HV-OVI absorbers. This locates the population of HV absorbers in the intergalactic space, possibly associated with an extended Galactic corona or Local Group medium.

We checked the validity of this method against our control-sample of LV-OVI absorbers, for which a nearby Galactic origin has been claimed[8,9,10,11,12]. The velocity signs of LV-OVI are distributed randomly in the LSR, but show a high degree of segregation in the GSR, with positive velocities segregated in the $0° \leq l \leq$

180° hemisphere, as expected for clouds co-rotating with the Galactic disk. Furthermore the amplitude of the average velocity vector of the LV-OVI increases monotonically from the LSR to the LGSR, from 5 km s$^{-1}$ up to 71 km s$^{-1}$, opposite to what is observed for the HV-OVI absorbers and again confirming the Galactic origin for the LV-OVI absorbers.

There are several pieces of evidence, both observational and theoretical, supporting the identification of the HV systems with tenuous and diffuse Warm-Hot gas filling our Local Group. Recent ''constrained'' hydrodynamical simulations for the formation of structures in our own Supercluster Environment[20,21] predict a major reservoir of baryons in the local environment in the form of WHIM filaments. The Local Group should be embedded in one of these filaments. The net motion of our Galaxy toward M 31, in the "local filament" (on average at rest in the LGSR), would produce an apparent bulk motion of the filament, towards our Galaxy, in the direction of M 31. This is similar to what is observed for the velocity field HV-OVI absorbers. The inverse of the mean GSR velocity vector of the HV-OVI absorbers (Table 1) has $<|v_{GSR}|^{HV}> = -64$ km s$^{-1}$, and Galactic coordinates of $l = 76°$, $b = -38°$, fairly close to M 31: $l = 121°$, $b = -22°$.

Estimates of the dynamical mass of our Local Group exceed the measured mass of its visible constituents by more than a factor of 2 (e.g. **4**). A total mass of $\sim 10^{12}$ M would be needed to stabilize, dynamically, the Local Group. From the combined analysis of the absorption features from highly ionized gas in the UV (HV-OVI) and X-ray (OVII, OVIII and NeIX) spectra of the blazar PKS 2155-304, Nicastro et al.[5] concluded that this absorber has to fill the intergalactic space surrounding the Galaxy. This system has a density of the order of $n_e \sim 4\text{-}6 \times 10^{-6}$ cm$^{-3}$, a column density of $N_H \sim 4.5 \times 10^{19}$ ([O/H]$_{0.3}$)$^{-1}$ cm$^{-2}$ (where [O/H]$_{0.3}$ is the O/H metallicity ratio in units of 0.3 times the solar value of $7.4 \times 10^{-4}$)[22] and so a linear size along the line of sight of $D \sim 2\text{-}4 \times$([O/H]$_{0.3}$)$^{-1}$ Mpc. If this particular line of sight is representative of the entire Local Group WHIM, then assuming a transverse size of $1 \times ([O/H]_{0.3})^{-1}$ Mpc, gives a total baryonic mass of 0.6-$2 \times ([O/H]_{0.3})^{-3} \times 10^{12}$ M$_\odot$, whose upper boundary would be sufficient to stabilize the Local Group.

Additional evidence supporting the Local-Group origin for the HV-OVI comes from the velocity distribution of the isolated Compact High Velocity Clouds of HI (CHVCs), studied in the 21cm HI emission line[23,24,25,26]. Based on statistical and theoretical arguments similar to those used here for the HV-OVI absorbers, it has been proposed that CHVCs originate at large (d > 100 kpc) characteristic distances[23,24,27]. Braun & Burton[25] estimated distances of 150 to 800 kpc for 10 of these objects, supporting a Local Group deployment. More recently De Heij & Burton[26] constructed kinematic models for the CHVCs and concluded that the majority of the CHVCs are at an average distance of 200 kpc from either our Galaxy or M 31. Sternberg, McKee and Wolfire[27] argued that the CHVCs may be viewed as dark-matter dominated "mini-halos" in orbit around the Galaxy (or M 31) at characteristic distances of 150 kpc. In their model a hot Galactic corona is actually required to pressure confine the HI gas inside the mini-halos.

Based on these findings, and on the dramatically different ionization state of the HV-OVI absorber presented here, and that of the isolated CHVCs, we suggest that CHVCs are the cold, condensed component of a multiphase local IGM in pressure equilibrium. The more diffuse and tenuous warm medium, possibly pervading the whole Local Group (and so spreading over distances of few Mpc), is responsible for the observed HV-OVI absorption, the X-ray OVII, OVIII and NeIX absorption systems, and provides the pressure needed to confine the cold phase found in the form of CHVCs. The hot plasma contains up to hundred times the total mass of the cold component[24]. The upper boundary of our baryonic mass estimate for the warm-hot phase suggests that this can account for up to 100% of the "missing baryons" in our Local Group.

Methodology

We used only FUSE data from the LiF mirror and the A1 detector segment, which have optimal efficiency in the waveband around the OVI line. The data were calibrated and cleaned following standard procedures[28]. The resolution of the final data is ~ 20 km s$^{-1}$, at 1032 Å.

We selected for our sample only spectra with signal-to-noise ratio (S/N) per resolution element of > 5 at 1032 Å since we were interested in the OVI absorption lines. This selection produced 54 different lines of sight, all with Galactic latitude $|b| > 20^o$ (because of the high extinction suffered by lines of sight crossing a significant portion of the Galactic disk). At this lowest quality spectra (S/N~5), absorption lines with equivalent width (EW) larger than ~ 200 mÅ are detected at > 3 σ, while the highest signal-to-noise spectra in our sample guarantee the detection of absorption lines down to EW ~ 20 mÅ.

For each z ~ 0 OVI absorption line spectrum we used the CIAO fitting engine "Sherpa"[29] to fit a power law continuum plus negative Gaussians to represent the absorption lines, and derived the best fit wavelengths and widths 1028.5 Å to 1034 Å.

We also looked for possible contamination by $H_2$ absorption lines from the galactic ISM, exploiting H2 templates computed using a non-LTE photodissociation front code[30,31]. The excitation temperatures of the $H_2$ rotational-vibrational level distribution result from formation pumping, UV absorption of diffuse UV starlight in the electronic transitions to the states B, C, B', and D followed by fluorescence, quadrupole cascading, collisional excitation and de-excitation. Templates have been computed for broad ranges of kinetic temperatures, densities, doppler parameters and factors of enhancing of the interstellar radiation field. Using these templates, we verified that, in the wavelength range of interest, spectra along lines of sight with equivalent hydrogen column densities lower than ~ $2\times10^{20}$ cm$^{-2}$ show only weak H$_2$ absorption. Spectra along lines of sight with $N_H > 2\times10^{20}$ cm$^{-2}$, instead, show usually strong (6-0) P(3) $\lambda$ 1031.19 Å and R(4) 1032.35 Å absorption. The P(3) $\lambda$ 1031.19 Å and R(4) 1032.35 Å lines occur at velocities of -214 and +123 km s$^{-1}$, and so may potentially contaminate HV-OVI lines. We found that about 30% of the negative HV-OVI lines and less than 20% of the positive HV-OVI lines had profiles contaminated by the P(3) $\lambda$ 1031.19 Å and R(4) 1032.35 Å lines respectively. For all spectra we used other strong H$_2$ lines in otherwise relatively featureless portions of the FUSE spectra, to determine the bulk velocity of the H$_2$ absorber and the strength the P(3) $\lambda$ 1031.19 Å and R(4) 1032.35 Å lines from the relative oscillator strength ratios, allowing us to disentangle the HV-OVI line from the contaminating H$_2$. In particular we used the (6-0) P(2) $\lambda$ 1028.10 Å, (6-0) R(3) $\lambda$ 1028.98 Å and (8-0) P(4) $\lambda$ 1012.26 Å lines, shortward of the OVI $\lambda$ 1031.926 Å line, and the (5-0) P(2) $\lambda$ 1040.36 Å, (5-0) R(3) $\lambda$ 1041.16 Å and (5-0) R(4) $\lambda$ 1044.54 Å lines, longward. Only in 2 cases could the presence of the HV-OVI line not be safely established, and so we considered those two lines of sight as free from HV-OVI absorption, at our detection threshold. In three more cases the presence of the HV-OVI line was clearly established, but its position and width only poorly constrained.

To identify HV-OVI lines with possible HV-HI counterparts, we cross-correlated the objects of our sample with catalogs of 21-cm HI lines[17,18]. We used very conservative searching criteria, and define a "coincidence" when Galactic coordinates and LSR velocities of the HV-OVI absorbers fall into the very broad ranges (typically tens of degree and about hundred km s$^{-1}$) listed in the HV-HI ctalogs. The very broad range in coordinates depends, not on the beam-size of the telescope with which 21-cm measures were performed, but on the entire size of the group or complex of HV-HI clouds to which a given cloud is supposed to belong. This search gave 5 "coincidences". For completeness we also cross-corelated our coincidences with those found by Sembach et al.[16] for their compilation of HV-OVI absorbers. This search left only one object. This object has been tentatively associated by Sembach et al. with the Magellanic Stream HV-HI complex, although not with any particular cloud. To be conservative we then add this object to the list of possible OVI-HI identification, and took these 6 objects off our sample for the purpose of the presented statistical analysis.

Acknowledgements

FN thanks P. Kaaret and G. C. Perola for fruitful discussions and useful comments. The authors thank Dr. W.B. Burton and an anonymous referee, for comments that helped improve the paper. This work has been partly supported by NASA-Chandra grants and the NASA-Chandra X-ray Center contract.

**Correspondence and requests for materials should be addressed to F.N. (fnicastro@cfa.harvard.edu)**

**Competing interests statement**

**The authors declare that they have no competing financial interests**

Table 1

| Reference Frame | $\langle v \rangle$ (km s$^{-1}$) | $\langle l \rangle$ (deg.) | $\langle b \rangle$ (deg.) |
|---|---|---|---|
| LSR | 121±17 | 277±12 | 33±27 |
| GSR | 64±12 | 284±19 | 38±45 |
| LGSR | 32±12 | 292±31 | 29±66 |

**The Amplitude of the Average HV-OVI Velocity Vector Minimizes in the LGSR.** Average (*v,l,b*) vectors, and associated uncertainties, for the HV-OVI absorbers, in the LSR, GSR and LGSR. Values are computed (a) projecting (*v,l,b*) onto three Cartesian axis (X,Y,Z), (b) averaging the components and transforming (<X>,<Y>,<Z>) back into (<*v*>,<*l*>,<*b*>), (c) computing the standard deviation of the Cartesian components, and propagating the errors in quadrature to obtain uncertainties on (<*v*>,<*l*>,<*b*>).

Figure 1

**The velocity range of HV-OVI absorbers greatly exceeds typical rotational velocities in the Galaxy.** Histogram of the HV-OVI (solid line) and LV-OVI (dashed line) velocity distributions in the LSR. Among the 54 objects of our sample, 45 (83 %) show at least one clear OVI absorption component at $z \sim 0$ at our detection thresholds (7 out of the remaining 9 objects have poor quality FUSE spectra, with detection threshold of EW $\geq 200$ mÅ). 4 show multiple LV and HV absorption. 38 lines of sight show LV-OVI absorption (70% of the sample), and 32 (59% of the sample) show HV-OVI components, with 22 objects showing both. Only 4 lines of sight show multiple LV or HV absorption.

Figure 2

**Velocity Segregation of the HV-OVI Absorbers.** Aitoff plots of the (a) LSR and (b) LGSR velocity distributions for the HV-OVI absorption systems. In both panels filled circles correspond to negative velocities, while open circles correspond to positive velocities. The size of the circles is proportional to the amplitude of the velocity. In the upper panel the open squares indicate the position of the barycenter of the Local Group (LG), M 31, and the Virgo cluster, while the star marks the barycenter of the distribution. Red circles mark HV-OVI absorbers tentatively associated with HV-HI clouds with known distance, which have therefore been excluded from our statistical analysis.

Figure 1

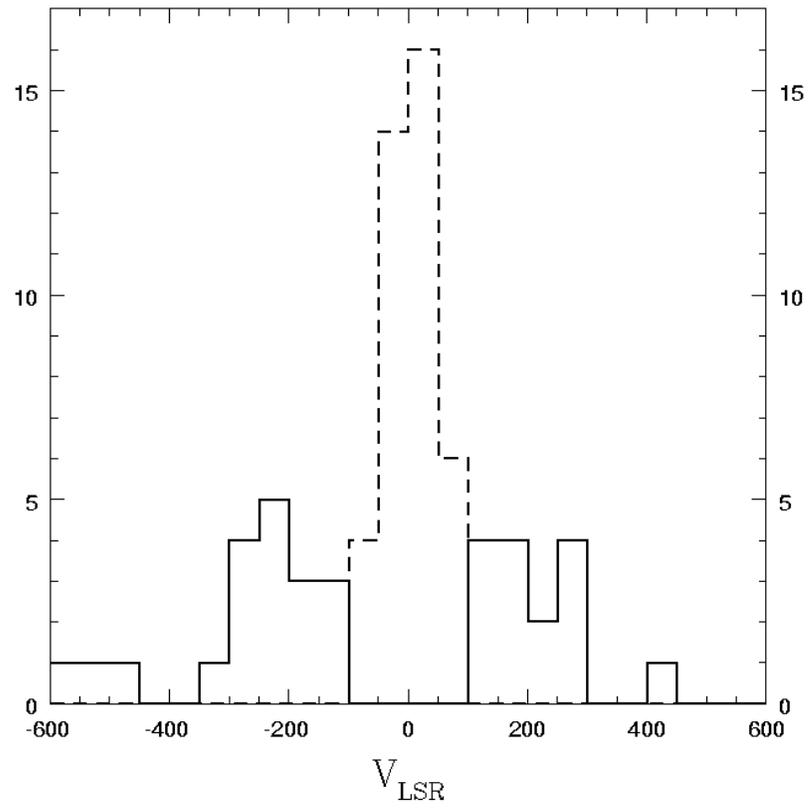

Figure 2

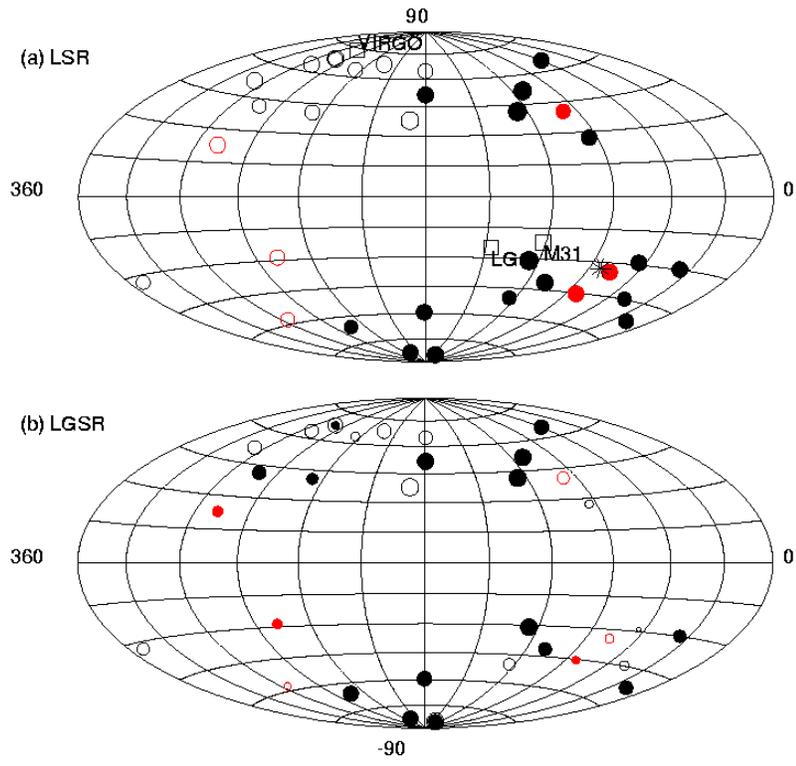